\documentclass[12pt,a4paper,dvips]{article}
\usepackage{a4p}
\usepackage{cite,mcite}
\usepackage{graphicx}
\usepackage{physics }
\usepackage{l3_title,ifthen}
\usepackage{rotating}
\journalname{Phys. Lett. B}

\usepackage{Lep}
\Lep{2}

\preprint{2000-120}
\date{11 September 2000}

\newlength{\capindent}
\newlength{\capwidth}
\newlength{\figwidth}
\newcommand{\icaption}[2][!*!,!]{\hspace*{\capindent}%
  \begin{minipage}{\capwidth}
    \ifthenelse{\equal{#1}{!*!,!}}%
      {\caption{#2}}%
      {\caption[#1]{#2}}
  \end{minipage}}
\newcommand{\pho}{\phantom{0}}

\newcommand{\qqll}{{\rm q \bar q\ell^+\ell^- }}
\newcommand{\eeqq}{{\rm e^+ e^- q \bar q }}
\newcommand{\mmqq}{{\rm \mu^+ \mu^-q \bar q }}
\newcommand{\ttqq}{{\rm \tau^+ \tau^- q \bar q }}
\newcommand{\eeto}{{\rm e^+e^- \rightarrow}}

\newcommand{\llnn}{{\rm \ell^+\ell^- \nu\bar\nu}}
\newcommand{\llll}{{\rm \ell^+\ell^-\ell^{\prime +}\ell^{\prime -} }} 
\newcommand{\qqnn}{{\rm q\bar q\nu\bar\nu}}
\newcommand{\qqqq}{{\rm q\bar qq^\prime\bar{q}^\prime}}

\begin{document}
\setlength{\unitlength}{1mm}
\begin{titlepage}
\title{Study of Z Boson Pair Production 
 in \boldmath{\epem} Interactions at \boldmath{$\sqrt{s}=192 - 202\GeV\,$}}
\author{The L3 Collaboration}
\begin{abstract}
The cross section for the production of Z boson pairs is measured
using the data   
collected by the L3 detector at LEP in 1999 in $\epem$ collisions at
centre-of-mass energies ranging from $192\GeV$ up to
$202\GeV$. Events in all the visible final states are 
selected, measuring the cross section of this process. The
special case of final states containing b quarks is also investigated.
All results are in agreement with the Standard
Model predictions.
\end{abstract}
\submitted
\end{titlepage}

\section{Introduction}                                         

The increase of the  LEP  centre-of-mass energy, $\sqrt{s}$,
beyond the Z pole has extended the range of the accessible physics
processes to include a sizable fraction
of four-fermion events. An important part of the four-fermion final states
emerges from the pair production of W or Z 
gauge bosons. 

The study of Z boson pair-production is of interest as it
offers a further test
of the Standard Model of the electroweak
interactions~\cite{sm_glashow}
in the neutral   
gauge boson sector. Moreover, this process 
constitutes a background in the search of the
Standard Model Higgs boson. In addition, Z-pair events allow the
investigation of possible triple neutral gauge boson couplings, ZZZ and
ZZ$\gamma$~\cite{hagiwara,gounaris2},
forbidden at tree level in the Standard  
Model. These events can also test new theories beyond the Standard Model
such as Supersymmetry~\cite{mele,gounaris2} or extra space
dimensions~\cite{agashe}. 

At the lowest order, Z pair-production proceeds via two $t$-channel
Feynman diagrams with an internal electron leg. Considering the Z decays into
fermions, this process is conventionally denoted as NC02, from the
acronym of the neutral-current production mechanism of the four-fermions
and the number of diagrams. A wider definition is used in
this letter, encompassing the regions of the full four-fermion phase
space compatible with the pair-production of Z bosons. Results in
the NC02 framework are also given. 

The experimental investigation of the Z pair-production is made difficult by
its rather low cross section, compared with competing two- and
four-fermion processes, that constitute large and sometimes irreducible
backgrounds.  
This process was observed at threshold by the L3 collaboration at 
$\sqrt{s} = 183\GeV$~\cite{zzl3183} and  studied later with a
higher statistical sample at $\sqrt{s}=189\GeV$~\cite{l3zz189}. 
Results from the other LEP collaborations were also
reported~\cite{opal,aleph}.
This letter
describes the extension of the L3 analyses to centre-of-mass energies
between $192\GeV$ and $202\GeV$. The measurement of the cross section
is presented together 
with other results that
include lower centre-of-mass energies. The measurement of the cross
section for the particular case of Z pair-production and decay into at
least a b quark pair is also discussed.

\section{Data and Monte Carlo Samples}

The data under investigation were collected in  1999 by the L3
detector~\cite{l3_00}
at four different centre-of-mass energies, 
$191.6\GeV$, $195.5\GeV$, $199.5\GeV$ and $201.7\GeV$ with
corresponding integrated luminosities of $29.7$\,pb$^{-1}$,
 $83.7$\,pb$^{-1}$, $82.8$\,pb$^{-1}$ and $37.0$\,pb$^{-1}$.
These energies are denoted as 192, 196, 200
and 202 hereafter.

The EXCALIBUR~\cite{exca} Monte Carlo is used to
generate events belonging to both the signal and the background
neutral-current four-fermion processes. Background from 
fermion-pair production is described making use of
PYTHIA~\cite{pythia} and KK2f~\cite{kk2f}
($\rm e^+ e^- \rightarrow q \overline q  (\gamma)$), 
KORALZ~\cite{koralz} and KK2f ($\rm e^+ e^- \rightarrow \mu^+
\mu^-  (\gamma)$ 
and $\rm e^+ e^- \rightarrow \tau^+ \tau^-  (\gamma)$)
and BHWIDE~\cite{bhwide} ($\rm e^+ e^-\rightarrow e^+ e^-  (\gamma)$).
Background from charged-current four-fermion processes
is generated with EXCALIBUR for $\rm e\nu_\e q\bar q'$ and
$\rm \ell^+\nu_\ell \ell^-\bar \nu_\ell$ with $\rm \ell=e,\mu,\tau$ and
KORALW~\cite{koralw} for W pair-production and decay in the final
states not covered by the simulations listed above.
Contributions from multi-peripheral processes
are modelled by
PHOJET~\cite{phojet} ($\rm e^+ e^- \rightarrow e^+ e^- q \overline q$)
and DIAG36\,\cite{diag36} ($\rm e^+ e^- \rightarrow e^+ e^-
\ell^+\ell^- $), in the quark and lepton low invariant mass region not
included in the samples generated with EXCALIBUR.

The L3 detector response is simulated using the GEANT
program~\cite{geant}, which takes into account the effects of energy loss,
multiple scattering and showering in the detector. 
Time dependent detector inefficiencies, as monitored during the data taking
period, are reproduced in these simulations.

The Z pair-production signal is defined as the subset of the full
four-fermion phase space satisfying the following
requirements~\cite{zzl3183,l3zz189}.  
First, the invariant mass of both
fermion pairs must be between $70\GeV$ and
$105\GeV$.  In the case in which
fermion pairs can originate from a charged-current process 
($\rm u\bar d d \bar u $, $\rm c\bar s s \bar c $   
and $\rm\nu_\ell\ell^+\bar{\nu_\ell} \ell^-$, with $\rm \ell=e,\mu,\tau$),
the masses of the fermion pairs which can also emerge 
from W decays  are required to be either below  $75\GeV$
or above $85\GeV$. Finally, events with electrons in the final state are
rejected if for any electron $|\cos{\theta_{\rm e}}| > 0.95$, where
$\theta_{\rm e}$  is the electron polar angle. 

The expected cross sections for the different final states are
computed imposing the requirements described above on a sample of
events generated  with EXCALIBUR, and found to be
0.79\,pb, 0.92\,pb, 1.00\,pb and 1.03\,pb at the four
centre-of-mass energies, in increasing order.
In this calculation  $\alpha_s=0.119$~\cite{pdg} 
is included for the QCD vertex corrections.
An uncertainty of $\pm 2\%$ is assigned to these predictions,
reflecting the differences between them and those obtained with the
GRC4F~\cite{grace} Monte Carlo  generator as well as the expected
accuracy of the treatment of initial state radiation.

The relative populations of the different channels, as obtained from
their corresponding cross sections, differ slightly from those of
the NC02 framework, derived from the Z branching ratios~\cite{pdg}.

The cross section for final states with at least one b quark pair 
is significantly smaller than the total cross section and all
centre-of-mass energies are hence combined. The corresponding predicted
cross section is 0.27\,pb with an  uncertainty of $\pm
2\%$.

\section{Event Selection}

All the visible final states of the Z-pair decay are investigated
with criteria similar to those used at
$\sqrt{s}=189\GeV$~\cite{l3zz189}, and modified to follow the
evolution of the signal topology, which
manifests a larger boost of 
the Z bosons with the higher $\sqrt{s}$.
All selections are based on the identification of two fermion pairs,
each with a mass close to the Z boson mass.

Electrons are recognised from energy depositions in the electromagnetic
calorimeter whose shower shape is compatible with that initiated by
an electron or a photon. According to the selection channel, a
track as reconstructed in the central tracker may be required to be
associated to this cluster.

Muons are reconstructed either from tracks in the muon spectrometer pointing
to the interaction vertex and in time with the event, or via only energy
depositions in the calorimeters consistent with a minimum ionising 
particle (MIP) which have a matching track in the central tracker.

Tau leptons are identified from their decay into electrons or muons or
as low-multiplicity jets with one or three associated tracks. 
A total unit charge is required.

Quarks manifest themselves with a high multiplicity 
of calorimetric clusters and charged tracks. These are grouped into
the required number of jets by
means of the DURHAM algorithm~\cite{durham}.

The hermeticity of the detector allows to reconstruct the
four-momentum of Z bosons decaying into neutrinos  by means of the
event  missing energy and the momentum imbalance.

The selection criteria specific for each final state are discussed
below, first for the channels containing hadrons, then for the purely
leptonic ones.

\subsection{The \boldmath{$\qqll$} Channel}

For each of the final states $\rm{q\bar{q}e^+e^-}$,
$\rm{q\bar{q}\mu^+\mu^-}$ and $\rm{q\bar{q}\tau^+\tau^-}$, a dedicated
selection is performed. A pair of leptons should be present in high
multiplicity events with visible energy and effective centre-of-mass
energy respectively in excess of 0.5$\sqrt{s}$ and 0.6$\sqrt{s}$. 
The effective centre-of-mass energy is the energy at which the
$\epem$ interaction takes place after the possible emission of  initial
state radiation photons. It is reconstructed taking into account both
photons observed in the detector and those collinear with the beam
axis~\cite{ff}. 
For the $\rm{q\bar{q}e^+e^-}$ selection, at least one electron is required
to have a matched track. No more  than one MIP is allowed in the
$\rm{q\bar{q}\mu^+\mu^-}$ selection. 

The $\rm{q\bar{q}\tau^+\tau^-}$ selection relies on 
both a particle-based and a jet-based approach. The
former is aimed to identify a pair of taus in the events while in the latter 
the event is constrained into four-jets.
Two of the jets must have less than four tracks and
are considered as the tau candidates. At least one of them
has also to  coincide with an identified tau.
The radiative   $\rm{q\bar{q}(\gamma)}$ background is further
suppressed by rejecting events
containing a photon of energy larger than $40\GeV$.

\begin{figure}[p]
\begin{center}
  \begin{tabular}{cc}
    \includegraphics[width=8.5truecm]{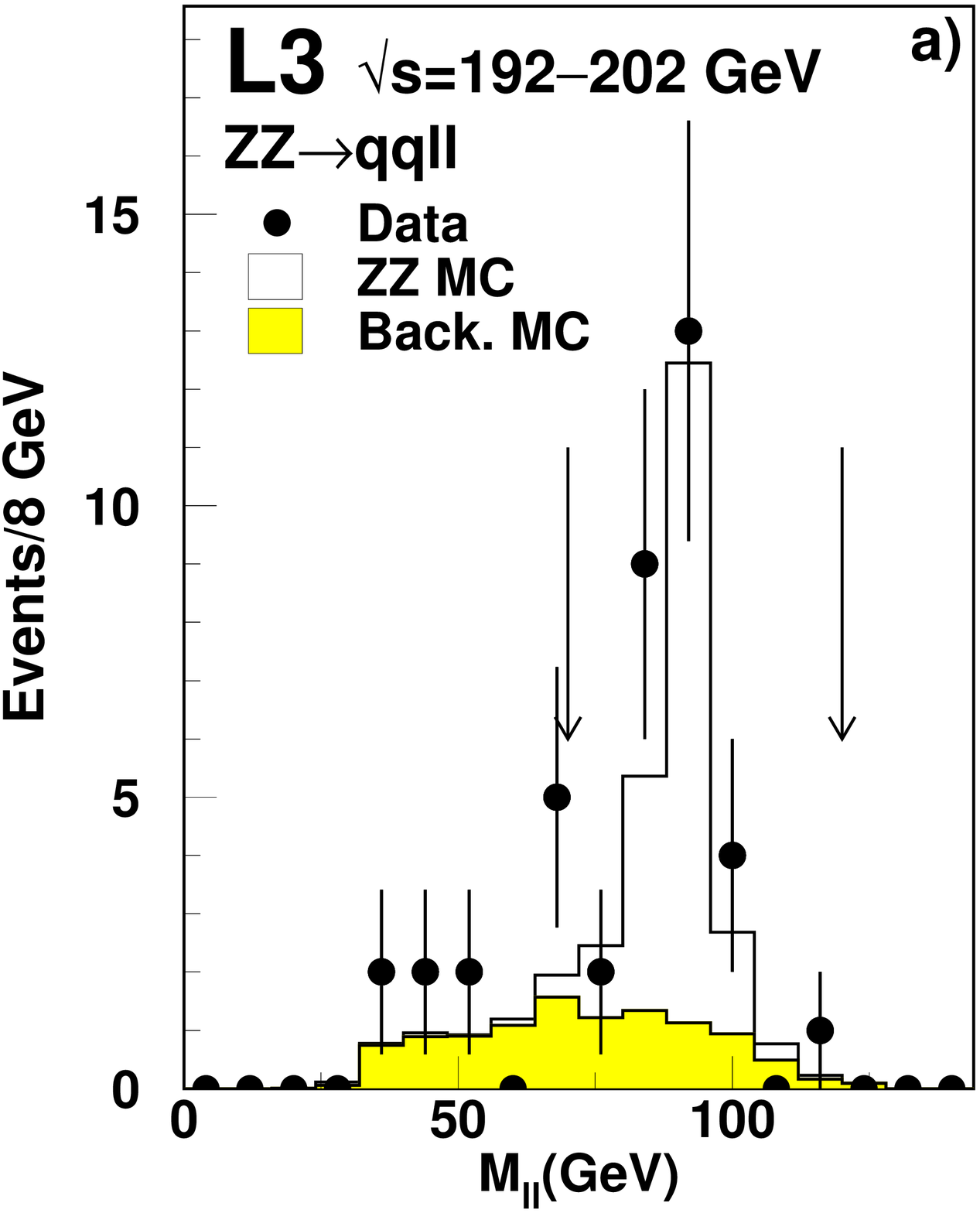} &
    \includegraphics[width=8.5truecm]{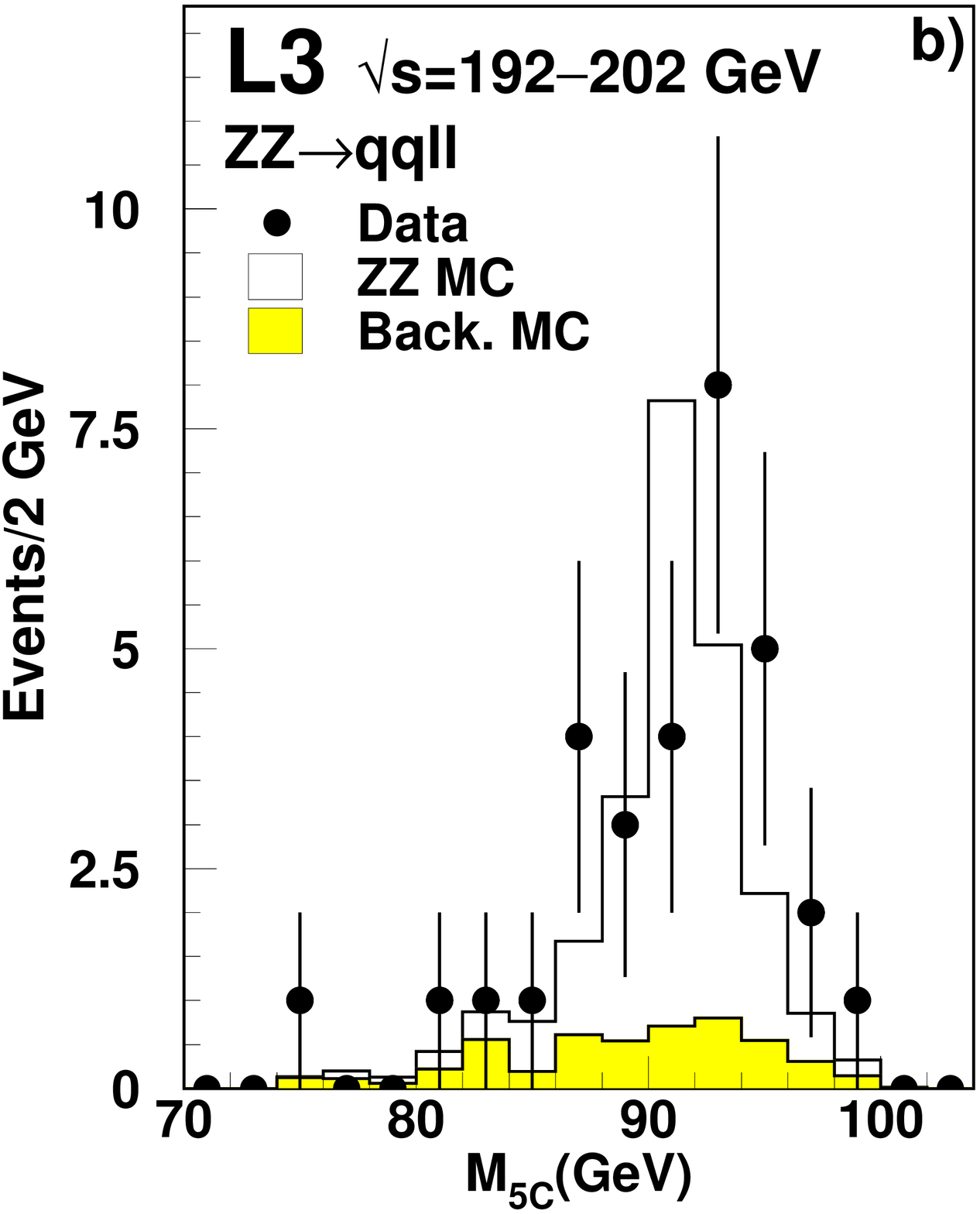} \\
    \includegraphics[width=8.5truecm]{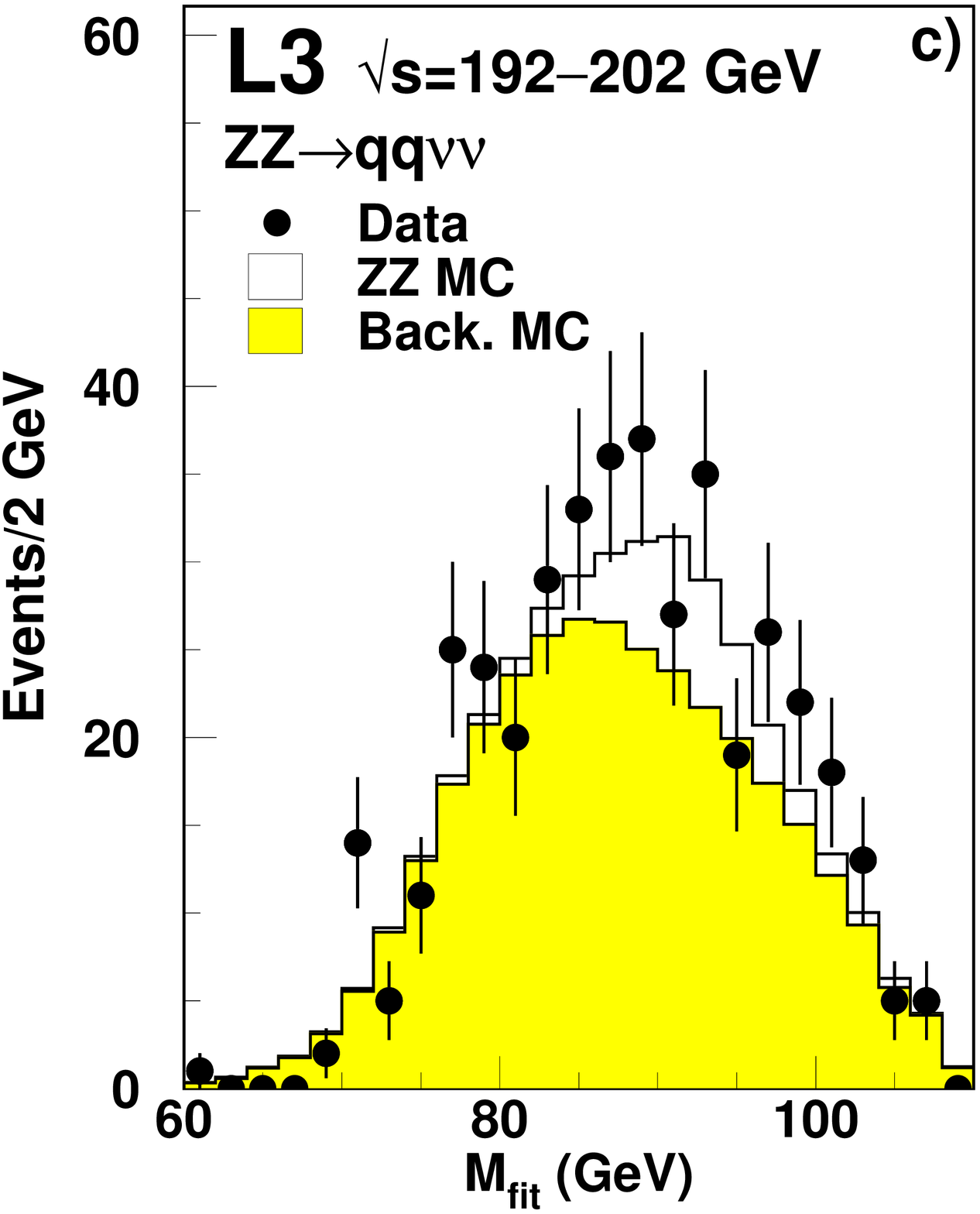} &
    \includegraphics[width=8.5truecm]{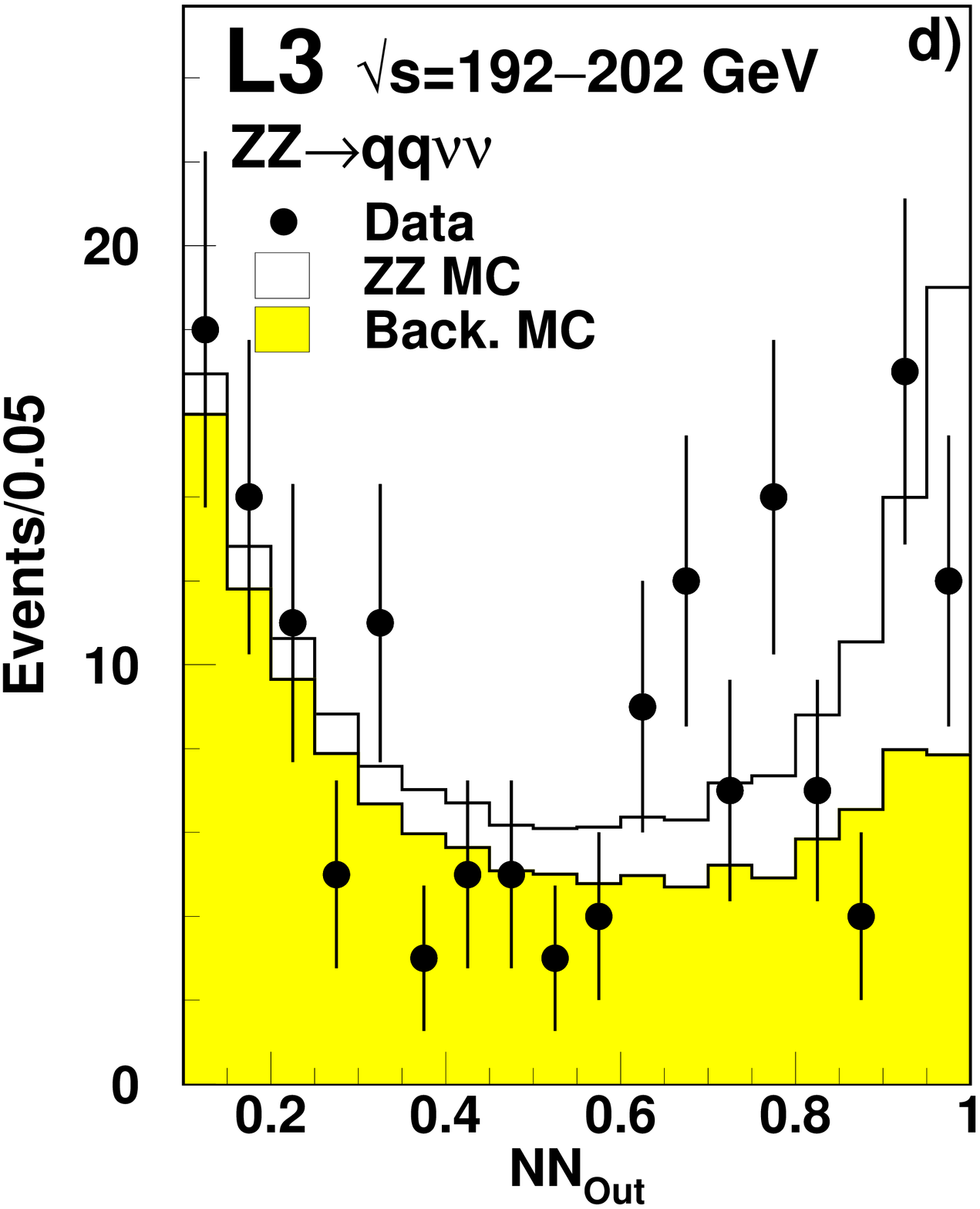} \\
  \end{tabular}
\caption{Distributions for data and Monte Carlo of: a) Invariant mass
    $M_{\ell\ell}$ of the lepton pair for the $\qqll$ final state before the
    application of the cuts  indicated by the arrows. b)
    The fit mass $M_{5C}$ of the $\qqll$ final state. c) The mass
    $M_{fit}$  of the
    hadronic system of the $\qqnn$ final state after a kinematic fit
    that imposes the Z mass to the event missing four-momentum. d)
    The output {\it NN}$_{Out}$ of the $\qqnn$  neural
    network.}
\label{fig:1}
\end{center}
\end{figure}

The topology of the pair-production of Z bosons is enforced by requiring
the lepton pair and the jet pair to have an opening angle of at least
$110^\circ$ for the electron and muon channels and $120^\circ$ for the
tau channel.
Moreover, the invariant mass of the jet-jet and the
lepton-lepton systems after performing a kinematic fit, which imposes
energy and momentum conservation, must be between $70\GeV$ and
$120\GeV$, as  depicted in Figure~\ref{fig:1}a for the lepton case.

The contribution from semileptonic decays of W pairs is reduced by requiring
the transverse missing momentum to be
lower than $0.2\sqrt{s}$ and
the visible energy in
the electron and muon channels  to be at least $0.85\sqrt{s}$, while
it has to be between 0.6$\sqrt{s}$ and 0.9$\sqrt{s}$ for the tau
selections. 

To reject the residual background from W pair-production and hadronic
events with gluon radiation, the events are subject to the DURHAM algorithm 
requiring $y_{34}$ to be greater than $0.001$ for the
electron and muon channels and $0.0025$ for the tau channel.  $y_{34}$
is the DURHAM resolution parameter for which events change from a
three-jet into a four-jet topology.

The kinematic fit is repeated on events that pass at least one of the
three selections described above, with the additional constraint of equal
invariant masses for the jet-jet and lepton-lepton systems. 
The distribution of this invariant mass, $ M_{5C}$, is shown in Figure~\ref{fig:1}b.
Table~1  summarises the efficiencies achieved by the different
selections. Tables~2 and~3 present respectively the total yield of the
selection and its breakdown into the different centre-of-mass energies.

\begin{table}[ht]
  \begin{center}
    \begin{tabular}{|c|c|c|c|c|}
      \hline
      & \multicolumn{4}{c|}{Selection} \\
      \cline{2-5}
      Final State & $\eeqq$ & $\mmqq$ & $\ttqq$ & Total \\
      \hline
      $\eeqq$     &  77.2\% & --      &  2.6\%  & 79.5\% \\
      $\mmqq$     &  --     & 53.7\%  &  6.2\%  & 59.2\% \\
      $\ttqq$     &   0.6\% &  0.2\%  & 28.2\%  & 28.7\% \\
      \hline
    \end{tabular}
    \caption{Efficiency of the   $\qqll$ selections and of their combination.} 
  \end{center}
\end{table}

\begin{table}[ht]
  \begin{center}
    \begin{tabular}{|c|r|r|r|c|}
      \hline
      Selection & Data & Signal MC & Background MC &
      Efficiency \\
      \hline   
      $\qqll$ & 31  & $18.8\pm 0.2$ & $   4.9\pm 0.4$ & $56.5\%$ \\  
      $\qqnn$ & 89  & $33.9\pm 0.2$ & $  57.7\pm 0.3$ & $55.4\%$ \\
      $\qqqq$ & 530 & $69.3\pm 0.3$ & $ 445.1\pm 3.3$ & $65.0\%$ \\
      $\llnn$ & 3   & $2.5 \pm 0.1$ & $   3.2\pm 0.1$ & $40.5\%$ \\ 
      $\llll$ & 3   & $1.3 \pm 0.0$ & $   1.0\pm 0.3$ & $39.4\%$ \\
      \hline
    \end{tabular}
    \caption{Data, signal and background Monte Carlo events selected
      by each analysis and their efficiency. The $\qqnn$ and $\qqqq$
      entries are  reported for 
      selection requirements of 0.5 and 0.2 on the neural network
      outputs, respectively. The
      $\llnn$ entries refer only to electrons and muons. The
      uncertainties shown are from Monte Carlo statistics.} 
  \end{center}
\end{table}

\begin{table}[ht]
  \begin{center}
        \begin{tabular}{|c|ccc|ccc|}
          \hline
          $\sqrt{s}$ &
          $N_D$ & $N_S$ & $N_B$ &
          $N_D$ & $N_S$ & $N_B$ \\
          \cline{2-7}
          $(\mathrm{Ge\kern -0.1em V})$           
          &
          \multicolumn{3}{c|}{$\qqll$} &
          \multicolumn{3}{c|}{$\qqnn$} \\
          \hline
          192  &  
          $ \pho2$ & $    2.0  \pm 0.1 $ & $    0.4  \pm 0.1$  &
          $ \pho3$ & $\pho3.5  \pm 0.1 $ & $\pho4.5  \pm 0.1$  \\
          196  &  
          $    13$ & $    6.7  \pm 0.1 $ & $    2.0  \pm 0.3$  &
          $    35$ & $   11.9  \pm 0.1 $ & $   18.6  \pm 0.1$  \\
          200  &  
          $    13$ & $    6.9  \pm 0.1 $ & $    1.8  \pm 0.3$  &
          $    35$ & $   12.6  \pm 0.1 $ & $   23.1  \pm 0.1$  \\
          202  &  
          $ \pho3$ & $    3.2  \pm 0.1 $ & $    0.7  \pm 0.1$  &
          $    16$ & $\pho5.9  \pm 0.1 $ & $   11.5  \pm 0.1$  \\
          \hline
          & 
          \multicolumn{3}{c|}{$\qqqq$} &
          \multicolumn{3}{c|}{$\llnn$} \\
          \hline 
          192  &  
          $\pho46$ & $\pho7.0  \pm 0.1 $ & $\pho49.0 \pm 1.7$  &
          $     0$ & $    0.28 \pm 0.02$ & $   0.44  \pm 0.02$ \\
          196  &  
          $   178$ & $   23.5  \pm 0.2 $ & $  155.1  \pm 1.9$  &
          $     1$ & $    0.82 \pm 0.06$ & $   0.88  \pm 0.04$ \\
          200  &  
          $   199$ & $   26.6  \pm 0.2 $ & $  162.8  \pm 1.7$  &
          $     2$ & $    0.92 \pm 0.04$ & $   1.38  \pm 0.12$ \\ 
          202  &  
          $   107$ & $   12.2  \pm 0.1 $ & $\pho78.2 \pm 1.2$  &
          $     0$ & $    0.45 \pm 0.03$ & $   0.49  \pm 0.03$ \\
          \hline
          &
          \multicolumn{3}{c|}{$\llll$}       &
          \multicolumn{3}{c|}{$\eeto\Zo\Zo$} \\
          \hline
          192  &  
          $     0$ & $    0.16 \pm 0.01$ & $ 0.24 \pm 0.04$    &
          $\pho51$ & $   12.9  \pm 0.1 $ & $\pho54.6 \pm 1.7$  \\
          196  &  
          $     1$ & $    0.50 \pm 0.02$ & $   0.42 \pm 0.15$  &
          $   228$ & $   43.4  \pm 0.2 $ & $  177.0  \pm 1.9$  \\
          200  &  
          $     2$ & $    0.44 \pm 0.02$ & $   0.30 \pm 0.09$  &
          $   251$ & $   47.4  \pm 0.2 $ & $  189.4  \pm 1.7$  \\
          202  &  
          $     0$ & $    0.21 \pm 0.01$ & $   0.09 \pm 0.19$  &
          $   126$ & $   22.0  \pm 0.2 $ & $\pho91.0 \pm 1.2$  \\
          \hline
        \end{tabular}
      \caption{Number of data ($N_D$), signal ($N_S$) and background
         ($N_B$)  Monte Carlo events selected  at the different
      centre-of-mass energies 
      in the separate final states and their sum. The 
      $\qqnn$ and $\qqqq$ 
      entries are reported for   
      selection requirements on the neural network outputs of 0.5 and
      0.2, respectively. Monte Carlo statistical uncertainties are given
      on the signal and  background expectations.} 
  \end{center}
\end{table}

\subsection{The \boldmath{$\qqnn$} Channel}

The selection of the $\qqnn$ channel proceeds from high multiplicity
events with an invariant mass in excess of $50 \GeV$. These
criteria deplete the total
data  sample of  purely leptonic two-fermion final states and 
products of two-photon interactions. Hadronic events from $\rm q
\bar{q} (\gamma)$ and W pair-production are then reduced by
requiring the invariant mass to be less than $130 \GeV$ and the mass
recoiling against the hadronic system to exceed $50 \GeV$.
Semileptonic decays of  W pairs are suppressed by rejecting events
with electrons or muons with energies above $20\GeV$.
The missing energy signature of a Z boson decaying into two neutrinos
is further exploited by requiring the transverse momentum  to
be above $5\GeV$, the 
energy deposition in the forward calorimeters to be below
$10\GeV$ and the missing momentum vector to point at least 
$16^\circ$ away from the beam axis. Moreover, the energy in a 25$^\circ$ azimuthal sector
around the missing energy direction, $E_{25}$, must not exceed
$30 \GeV$.

The selection requirements described above select 407 events in the full data
sample. The  Monte Carlo
expectations are 45 events for the signal and 339 for the background,
mainly accounted for by  charged-current
four-fermion processes. 

An artificial neural network is then designed
to further discriminate Z pair events from background. It is based on
event shape 
variables that differentiate the two-jet from the three-jet topology, on the
sum of invariant and missing masses, on the masses of the two jets
into which the event can be forced, the
total missing momentum and $E_{25}$. A constrained
fit is applied to the hadronic system in the hypothesis that the missing
energy and momenta are due to a Z boson. The resulting mass $
M_{fit}$, presented in Figure~\ref{fig:1}c, is also used in the
neural network.
The  output {\it NN}$_{Out}$ of the neural network 
is presented in Figure~\ref{fig:1}d.
The efficiency and the results of this selection are summarised in
Table~2 and detailed in Table~3 for the different centre-of-mass
energies for a benchmark cut of 0.5 on {\it NN}$_{Out}$.

\subsection{The \boldmath{$\qqqq$} Channel}

The $\qqqq$ channel is investigated by first selecting
high-multiplicity events with  a
visible energy between 0.6$\sqrt{s}$ and  
1.4$\sqrt{s}$, parallel and perpendicular imbalances below 0.3$\sqrt{s}$
and no identified electron, muon or photon with energy above
$65\GeV$.
The events are  forced to four jets and then
subjected to a constrained fit which rescales the jets to balance
momentum while imposing energy conservation, 
greatly reducing the dependence on the calorimeter energy scale.

Such hadronic events are copiously produced in
QCD processes and W-pair production. Two artificial neural networks
are sequentially 
constructed to isolate the Z pair signal and  reject these two
backgrounds.
The first neural network~\cite{ww} 
helps in selecting the signal from the QCD background. After a cut at
0.65 on the output 
of this neural network, displayed in Figure~\ref{fig:2}a,
the content in W and Z pair production events is enhanced.

A second neural network is then built to distinguish Z pairs from W pairs.
It relies on  
the reconstructed di-jet mass,  the maximum 
and minimum jet energy, the average number of charged tracks
per jet and the di-jet mass difference. 

Almost 40\% of the events generated in the $\qqqq$ channel and
satisfying the signal 
definition contain at least a b quark pair, while the b-content in W
pair events is negligible. A
b-tag discriminant~\cite{h183}, is then added to the
network to further discriminate Z pair  from W pair events.

Figure~\ref{fig:2}b displays the output of this network after  the W pair 
enriched region below 0.2 is discarded. Events compatible with Z
pair-production preferentially populate the region between 0.6 and
0.8 if their content in  b quarks is low and lie above otherwise.
The performances of this analysis are summarised in Tables~2 and~3.

These results are confirmed by
a simpler cut-based analysis that relies on the signature of the
different boost of   
Z and W pairs as retained in the two di-jet opening angles, the di-jet mass
difference and the di-jet mean mass. Another study mainly aimed at the
rejection of the QCD background and the simultaneous selection of
four-jet W
and Z pair events also yields compatible results. Both these analyses are
affected by a lower purity which follows from the absence of a b-tag.

\subsection{The \boldmath{$\llnn$} Channel}

A pair of identified electrons or muons constitutes the core of the $\llnn$
selection. Tracks are not required for the
electron identification and MIPs are not considered as muon
candidates. The lepton pair must be consistent with a Z
boson, with an
invariant mass, $M_{\ell\ell}$, between $80\GeV$ and $100\GeV$. The
recoil mass, $M_{rec}$, is required to lie in the same interval to
enforce the signature of the second Z decaying into two neutrinos.

Fermion-pair events are rejected by requiring the lepton pair to be
acoplanar and to have a visible energy compatible with the signal
hypothesis. Moreover, the missing 
momentum vector must point away from the beam line.

The background from
other resonant and non-resonant four-fermion processes is reduced by
performing  a kinematic fit which imposes the Z mass to the
visible pair of leptons. The recoil mass $M_{rec}^{fit}$ is
recalculated and required to be compatible with the Z mass.

The distribution of the sum of $M_{\ell\ell}$ and $M_{rec}$ is
expected to peak around twice the Z mass 
for signal events, and is presented in Figure~\ref{fig:2}c. The efficiency of
the selection is reported in Table~2, 
which also lists the total number of selected and expected events,
detailed in Table~3 for the different centre-of-mass energies.
No contribution from the $\tau^+\tau^-\nu\bar{\nu}$
signal is expected after this selection. The dominant background
arises from charged-current four-fermion processes.

\subsection{The \boldmath{$\llll$} Channel}

To achieve a high efficiency, the selection of the $\llll$ channel
starts from four or more loosely identified
leptons in low-multiplicity events and concentrates on the kinematic
properties of just a pair of 
them. Electrons with or without a matched track, muons and taus
are accepted in the first   
stage, provided
their energy exceeds $3\GeV$. If more than four leptons are present, the four  most compatible with energy and momentum
conservation are selected. 

The event must contain at least an electron or a muon pair. To form
such pairs at least one electron
should have a matched track and no MIPs are considered as muons. In
the case of multiple 
choices, the pair with the invariant mass  $M_{\ell\ell}$
closest to the Z mass is studied.
Both $M_{\ell\ell}$ and the recoil mass $M_{rec}$ to this
selected lepton pair are 
required to be in the range between $70\GeV$ and $105\GeV$. 

Selection criteria on the energy of the most energetic electromagnetic
cluster of the event and  the
acoplanarity and  acollinearity of  the lepton pair
reject the residual Bhabha  and  
radiative fermion pair-production backgrounds.

The data and Monte Carlo distributions for the sum of $M_{\ell\ell}$ 
and $M_{rec}$,
the most discriminating variable between signal and background,
are displayed in Figure~\ref{fig:2}d. The yield of the selection for the total
sample, and the separate energies are respectively given in
Tables~2 and~3. The background is mainly constituted by
non-resonant neutral-current four-fermion events.

\section{Results}

\subsection{Measurement of the ZZ Cross Section}

\begin{figure}[p]
\begin{center}
\begin{tabular}{cc}
\mbox{\includegraphics[width=8.5truecm]{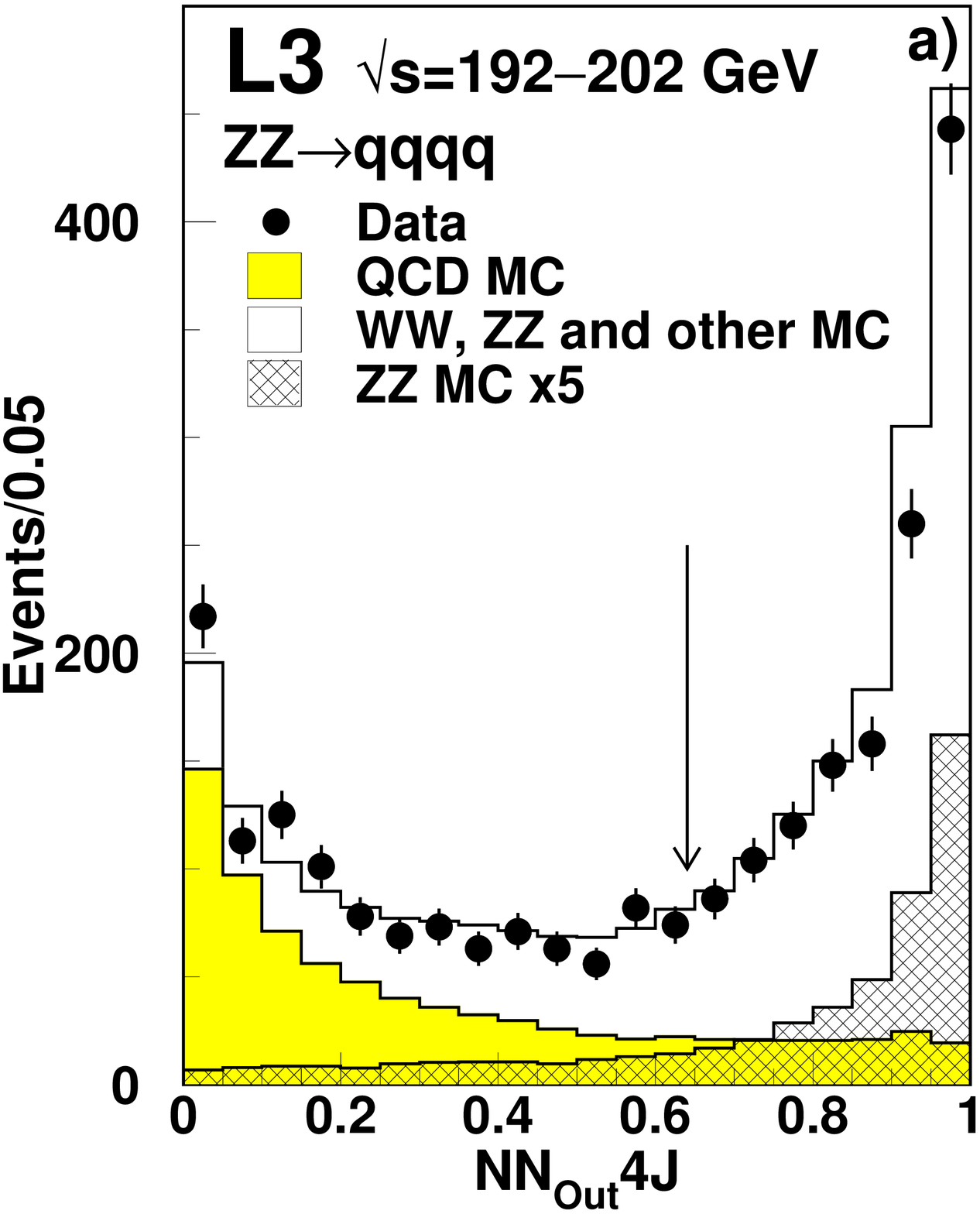}} &
\mbox{\includegraphics[width=8.5truecm]{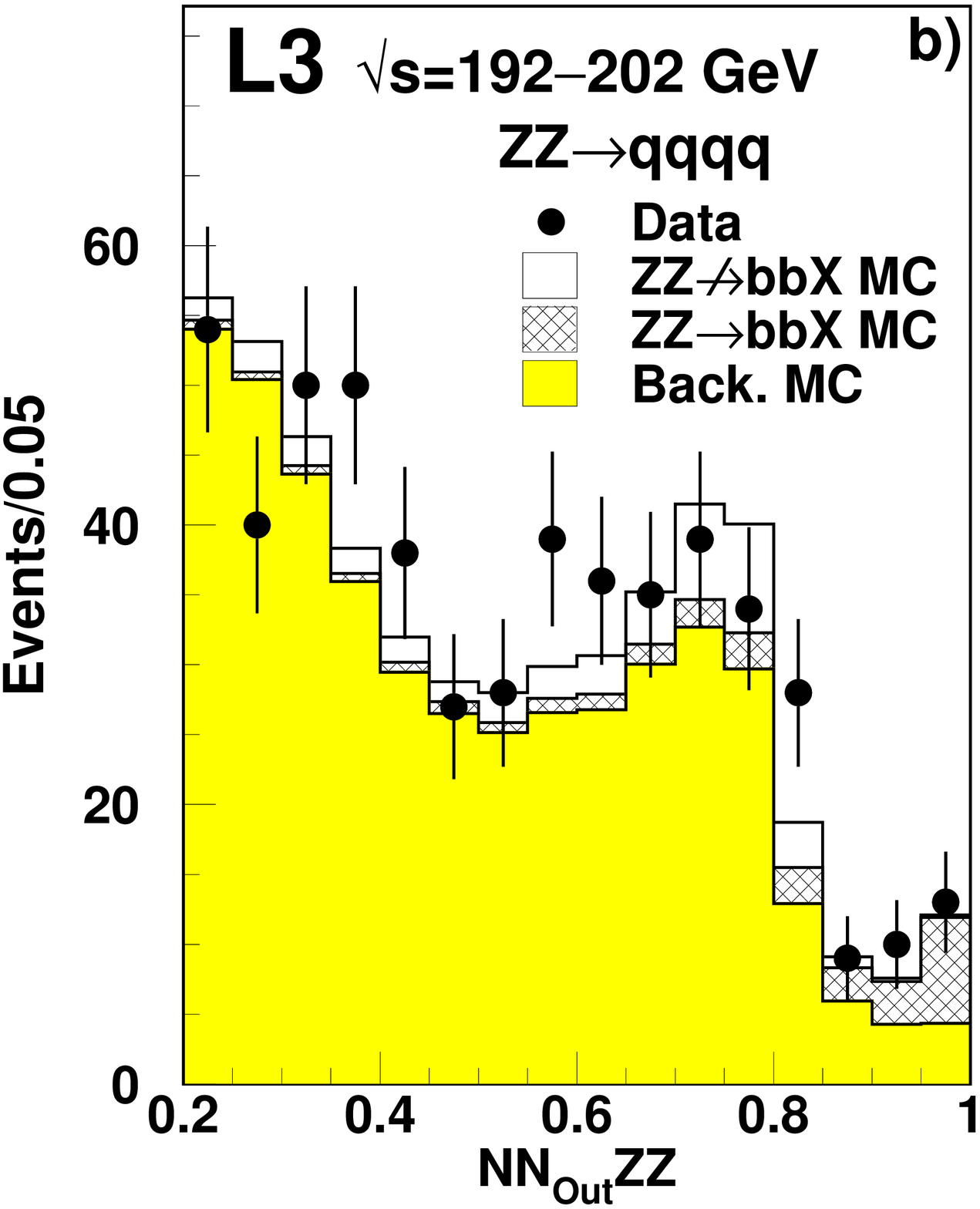}} \\
\mbox{\includegraphics[width=8.5truecm]{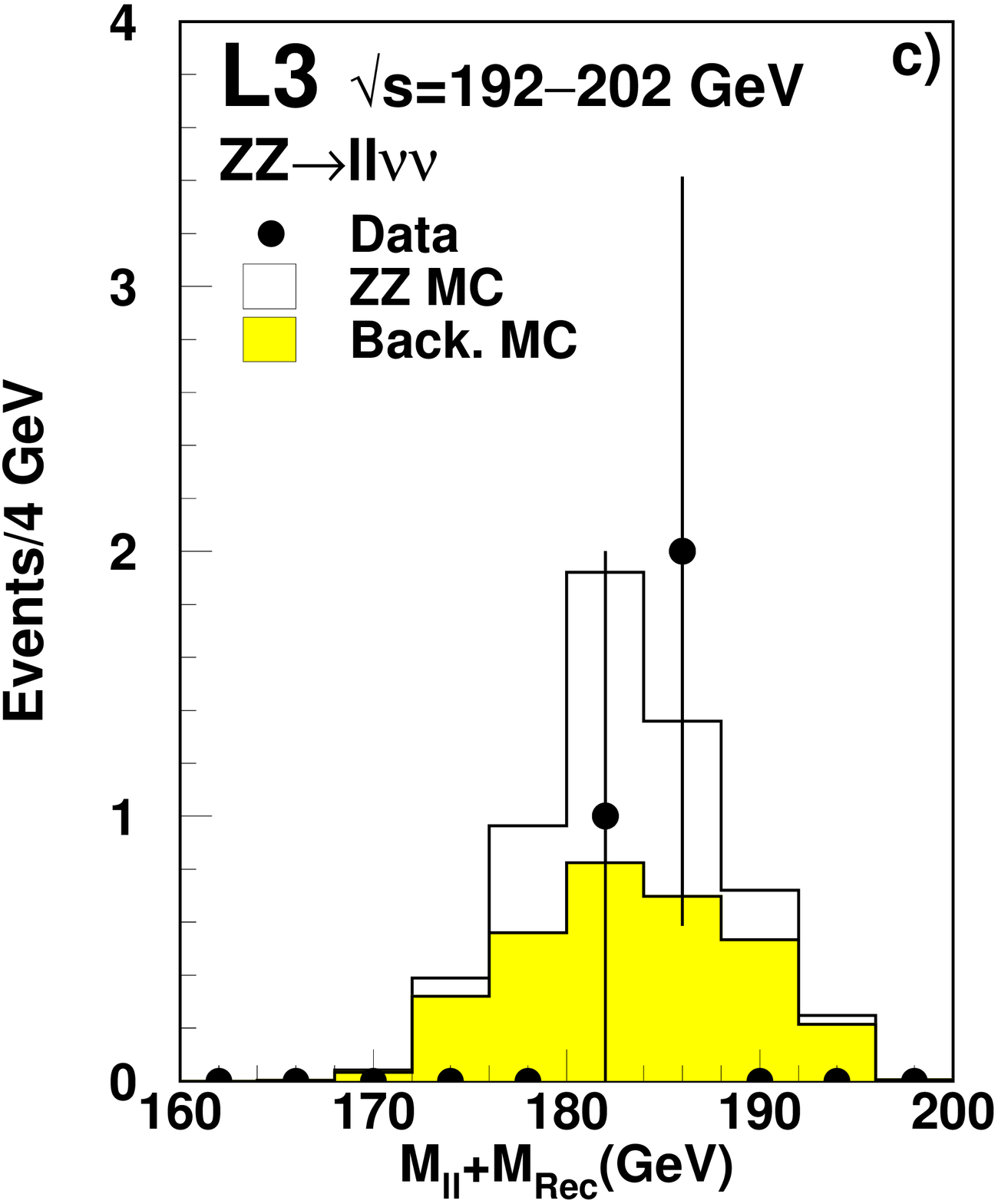}} &
\mbox{\includegraphics[width=8.5truecm]{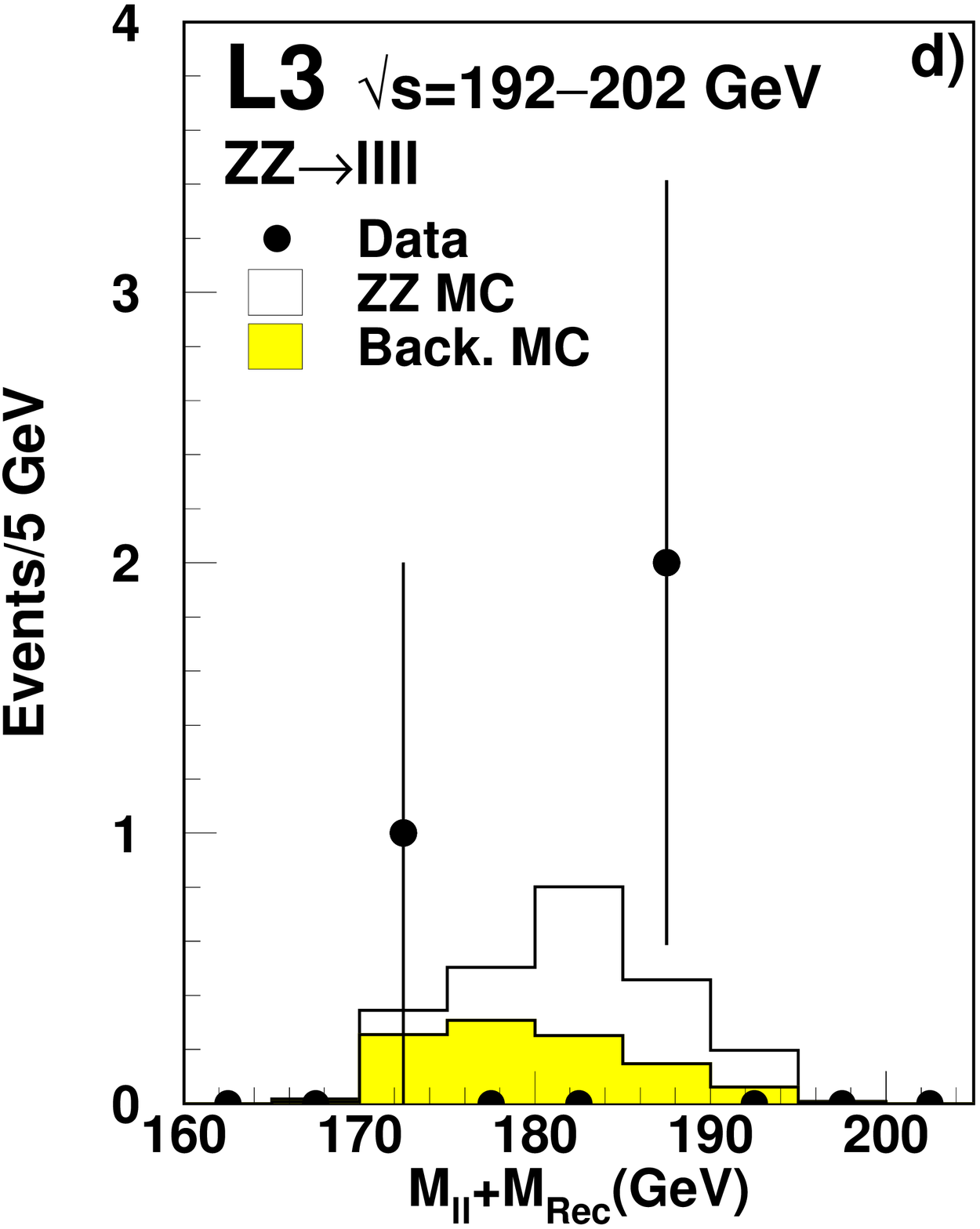}} \\
\end{tabular}
\caption{Distributions for the $\qqqq$ selection of the outputs  a) 
  {\it NN}$_{Out}${\it 4J} of the  first
  neural network; the ZZ signal is
  superimposed with a cross section five times larger than the
  predicted one and the arrow shows the cut, b)  {\it NN}$_{Out}${\it ZZ} of
  the final neural network;
  signal expectations for events with no or at least one b quark pair
  are presented separately.
  Distributions of the sum of the visible and recoil masses for  c) the  
  $\llnn$ and d) the $\llll$ selections. Data and Monte
  Carlo are shown.}
\label{fig:2}
\end{center}
\end{figure}

The distributions of the variables presented in
Figures~\ref{fig:1}b,~\ref{fig:1}d,~\ref{fig:2}b,~\ref{fig:2}c
and~\ref{fig:2}d are separated into each centre-of-mass energy and
are then 
fit to determine the cross section of the individual channels. 

A probability density function  is built
from the observed number of events in each of the bins of the
distribution as a function of  the signal cross section, fixing the
background expectations. A flat positive  
distribution for its value is assumed. 
If a zero value of the cross section is contained in a $\pm 34\%$
confidence interval around the maximum of the probability density
function, then a 95\% confidence level 
upper limit is calculated. This maximum is otherwise quoted as the
measurement, adopting  this interval as the corresponding
statistical uncertainty.  

Table~4 lists the results of all these fits together with the Standard
Model predictions.  Assuming these
predictions as the relative weights of the different channels, the 
ZZ cross section  for each centre-of-mass energy can be calculated
from a simultaneous fit to the five channels. The results of this
fit are also presented in Table~4. All the measured cross sections agree with
the Standard Model predictions.
In the calculation of the cross section, the effect of the cross talk
between the separate channels is found to be negligible.

\begin{table}[ht]
  \begin{center}
        \begin{tabular}{|c|cc|cc|cc|}
          \hline
          $\sqrt{s}$ &
          $\sigma^{fit}$\,(pb)& $\sigma^{th}$\,(pb) &
          $\sigma^{fit}$\,(pb)& $\sigma^{th}$\,(pb) &
          $\sigma^{fit}$\,(pb)& $\sigma^{th}$\,(pb) \\
          \cline{2-7}
           ($\mathrm{Ge\kern -0.1em V}$) & 
          \multicolumn{2}{c|}{$\qqll$} &
          \multicolumn{2}{c|}{$\qqnn$} &
          \multicolumn{2}{c|}{$\qqqq$} \\
          \hline 
          192  &  
          $<0.36$ & $0.12$ & 
          $<0.28$ & $0.22$ & 
          $<0.73$ & $0.38$ \\ 
          196  &  
          $0.20 \pm 0.07$ & $0.14$ & 
          $0.25 \pm 0.11$ & $0.25$ & 
          $0.63 \pm 0.20$ &  $0.44$ \\ 
          200  &  
          $0.22 \pm 0.08$ & $0.15$ &
          $0.25 \pm 0.12$ & $0.28$ & 
          $0.60 \pm 0.20 $ & $0.48$ \\
          202  &  
          $<0.32$ & $0.15$ & 
          $0.16 \pm 0.15$ & $0.29$ & 
          $0.84 \pm 0.33$ & $0.49$ \\
          \hline
          &
          \multicolumn{2}{c|}{$\llnn$} &
          \multicolumn{2}{c|}{$\llll$} &
          \multicolumn{2}{c|}{$\eeto\Zo\Zo$} \\
          \hline
          192  &  
          $<0.26$ & $0.03$ & 
          $<0.11$ & $0.01$ &
          $0.29 \pm 0.22$ & $0.79$ \\
          196  &  
          $<0.19$ & $0.04$ & 
          $<0.11$ & $0.01$ &
          $1.17 \pm 0.24$ & $0.92$ \\
          200  & 
          $<0.24$ & $0.04$ & 
          $0.06 \pm 0.04$ & $0.01$ &
          $1.25 \pm 0.25$ & $1.00$ \\
          202  &  
          $<0.26$ & $0.04$ & 
          $<0.13$ & $0.01$ &
          $0.93 \pm 0.38$ & $1.03$ \\
          \hline
        \end{tabular}
    \caption{Results, $\sigma^{fit}$, of the individual and global cross section
          fits for the different centre-of-mass energies. The
          corresponding theory predictions, $\sigma^{th}$, are also given.
           Limits are at 95\% confidence level.}
  \end{center}
\end{table}

\begin{figure}[p]
\begin{center}
    \begin{tabular}{c}
      \mbox{\rotatebox{90}{
          \includegraphics[width=.72\figwidth]{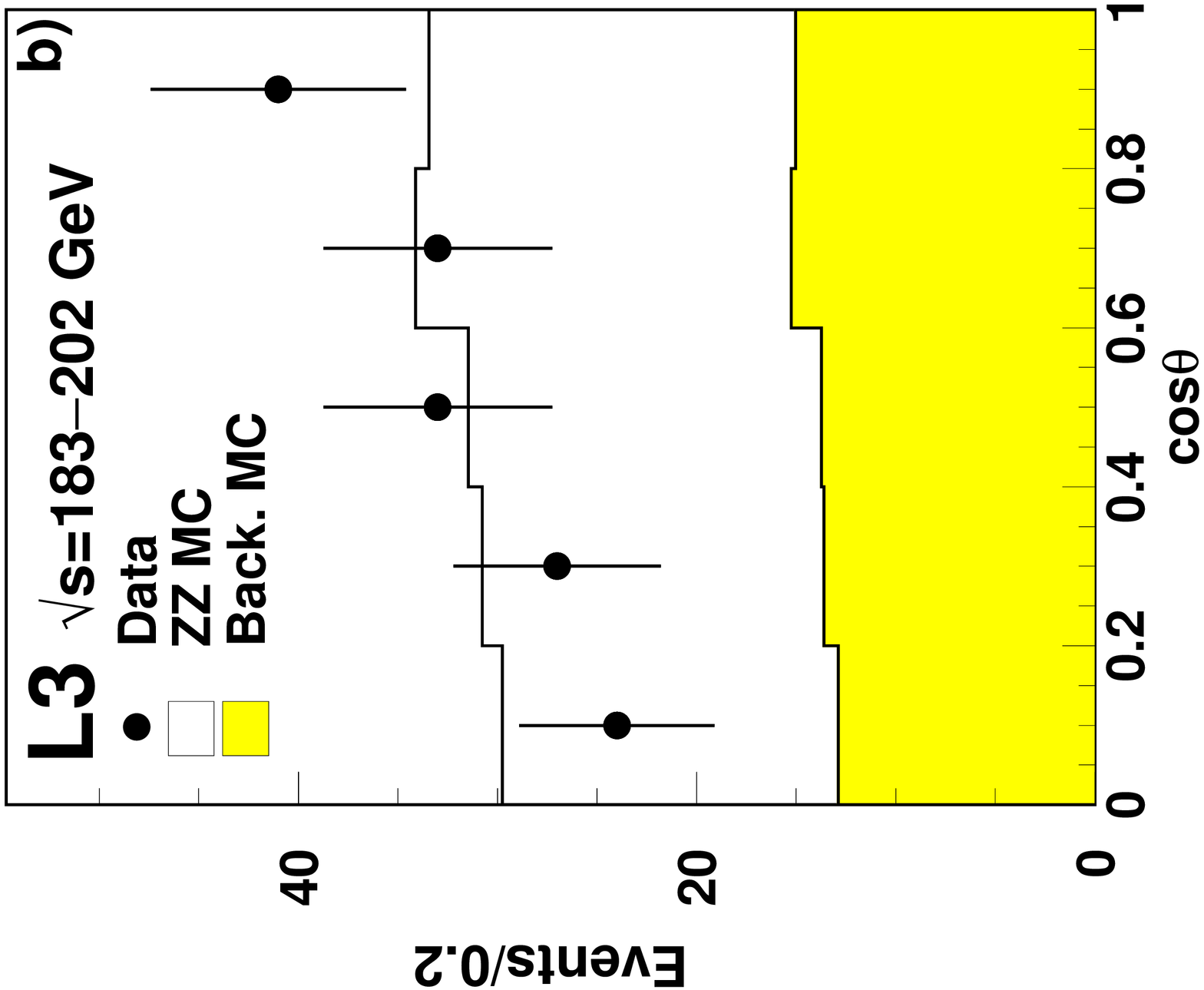}}} \\
      \mbox{\rotatebox{90}{
          \includegraphics[width=.72\figwidth]{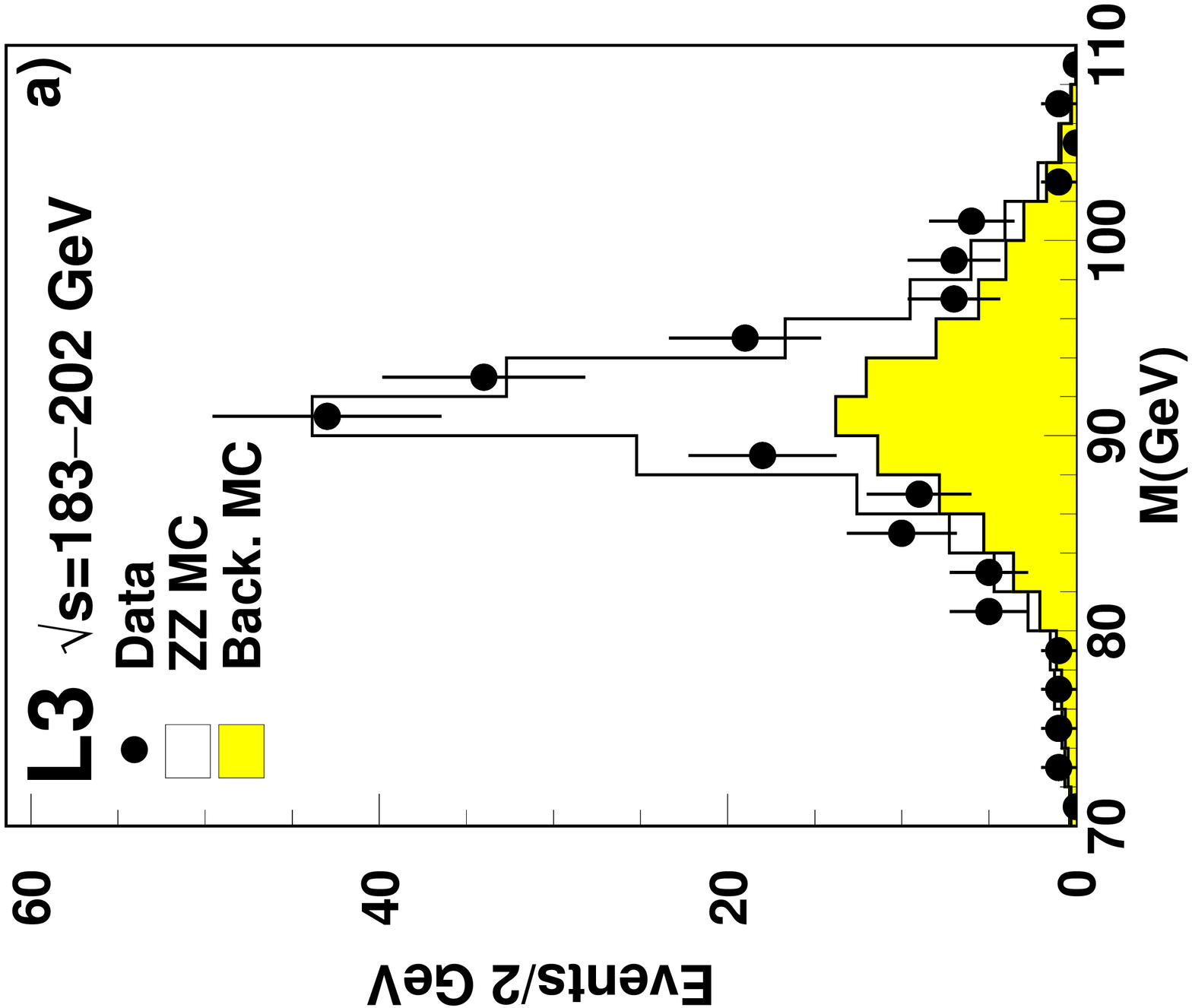}}} \\
    \end{tabular}
\caption{Distributions in data and Monte Carlo at all the LEP 
centre-of-mass energies  above the $\Zo$ pair production threshold
of a) the reconstructed  mass $M$ and b) the cosine of the
production angle $\theta$. Cuts on the $\qqnn$ and $\qqqq$ neural
network outputs are applied as 0.8 and 0.85, respectively.} 
\label{fig:3}
\end{center}
\end{figure}

Figure~\ref{fig:3}a presents the distribution of the reconstructed
mass, $M$, for all the selected events, including those collected at
lower centre-of-mass 
energies~\cite{zzl3183,l3zz189}. A cut on the $\qqnn$ and $\qqqq$
neural network outputs at 0.8 and 0.85 is applied, respectively.
For the $\qqll$ and the $\qqnn$ channels,  $M$ corresponds to ${
  M_{5C}}$ and ${ M_{fit}}$, respectively. The 
average of the two di-jet masses is considered for the $\qqqq$
channel while for both the $\llnn$ and $\llll$ channels the average of
$M_{\ell\ell}$ and  $M_{rec}$ is used.
A fit to the distribution of $M$ is performed in terms of the ratio
$R_{\rm ZZ}$ between the observed events and the  
predictions from Z pair-production and yields:
\begin{displaymath}
 R_{\rm ZZ} = 0.94 \pm 0.14\pm 0.06,
\end{displaymath}
in agreement with the Standard Model. The first uncertainty is
statistical and the second systematic, discussed in
References~\citen{zzl3183} and~\citen{l3zz189} and below.
 The cosine of the observed production polar angle $\theta$ is
 presented in Figure~\ref{fig:3}b for the same selected events. 

\subsection{Study of Systematic Uncertainties}

Systematic uncertainties on the $\eeto\Zo\Zo$ cross section can be
divided into sources correlated and uncorrelated among the channels.
Their effects are estimated using the full 1999 data sample, and then
propagated to the measurements performed at the different
centre-of-mass energies. 

The main sources of correlated systematic uncertainty are the background 
cross sections and the energy scale of 
the detector.  As they modify the shapes of the fit
distributions, their effect is evaluated
performing a new fit to calculate the cross section once their
values are modified as  listed in Table~5.  Possible non-linearity
effects for the energy scale are investigated.
The effect of the uncertainty of the LEP beam energy is negligible.

An  uncertainty  of 2\% is attributed to the measured
cross section to take into account the difference of the assumed relative
weights of the different channels, given by the EXCALIBUR calculation,
with respect to those
obtained with GRC4F, and to
parametrise other uncertainties related  to their calculation.

Some sources of systematic uncertainty are uncorrelated among the
channels but  modify the shape of the output of the final neural
network of the $\qqqq$ selection. These are the 
jet resolution,  the charged track multiplicity and the b-tag, and
their effect is presented in Table~5. The jet
resolution includes a variation of $\pm 2^\circ$ on the jet direction
and a smearing of $y_{34}$. A variation of the b-tag
discriminant of $\pm 2\%$ models possible systematic effects and
includes uncertainties in the Monte Carlo description of b-hadron jets.

\begin{table}[ht]
  \begin{center}
    \begin{tabular}{|c|c|r|r|}
      \hline
      Systematic Source & Variation & $\delta\sigma_{\rm ZZ}$ (\%) &
      $\delta\sigma_{\rm ZZ\ra b\bar{b}X}$   (\%)  \\
      \hline
      \rule{0pt}{12pt}Correlated sources                       &          &                  &   \\
      \hline
      \rule{0pt}{12pt}WW cross section              & 2\%      &$   2.4$&$   2.6$ \\
      \rule{0pt}{12pt}Four-jet rate                & 5\%      &$   2.1$&$   3.0$ \\
      \rule{0pt}{12pt}W$\rm{e}\nu$ cross section    & 10\%     &$   1.3$&$ < 0.1$ \\
      \rule{0pt}{12pt}Four-fermion cross section   & 5\%      &$   0.3$&$   2.6$ \\
      \rule{0pt}{12pt}Energy scale                  & 2\%      &$   3.6$&$   3.9$ \\
      \rule{0pt}{12pt}Lep energy                    &$40\,\MeV$&$ < 0.1$&$ < 0.1$ \\
      \rule{0pt}{12pt}Theory predictions            & 2\%      &$   2.0$&$   2.0$ \\
      \hline                                                                 
      \rule{0pt}{12pt}Uncorrelated sources                       &      &              &   \\
      \hline                                                                 
      \rule{0pt}{12pt}Jet resolution  ($\qqqq$)     & see text &$   0.3$&$   0.4$ \\
      \rule{0pt}{12pt}Charge multiplicity ($\qqqq$) & 1\%      &$   2.0$&$   2.4$ \\
      \rule{0pt}{12pt}B-tag  ($\qqqq$)             & 2\%      &$   1.6$&$   7.5$ \\
      \rule{0pt}{12pt}Monte Carlo statistics        & see text &$   3.9$&$   3.3$ \\
      \rule{0pt}{12pt}Simulation/Lepton Id          & see text &$   2.5$&$   1.4$ \\
      \hline                                                                 
      \rule{0pt}{12pt}Total                         &          &$   7.5$&$  10.8$ \\
      \hline
    \end{tabular}
    \caption{Systematic uncertainties on  $\sigma_{\rm ZZ}$ and 
      $\sigma_{\rm ZZ\ra b\bar{b}X}$. The total systematic uncertainty
      is the sum in quadrature of the different contributions.}
  \end{center}
\end{table}

\begin{table}[ht]
  \begin{center}
    \begin{tabular}{|c|c|c|c|c|c|}
      \hline
                               & $\qqll$ & $\qqnn$  & $\qqqq$ & $\llnn$ & $\llll$\\
      \hline                                                                   
      Signal MC statistics  ($\sigma_{\rm ZZ}$) 
                               &      1.1\%  &  0.4\%   &  0.4\%  &       3.2\% &   \pho2.3\% \\
      Background MC statistics ($\sigma_{\rm ZZ}$) 
                               &      8.1\%  &  0.3\%   &  0.7\%  &       4.1\% &      24.8\% \\
      \hline                                                                   
      Signal MC statistics   ($\sigma_{\rm ZZ\ra b\bar{b}X}$)
                               &      1.5\%  &  1.3\%   &  0.7\%  &          -- &         --  \\
      Background MC statistics  ($\sigma_{\rm ZZ\ra b\bar{b}X}$)
                               &      8.1\%  &  0.6\%   &  1.3\%  &          -- &         --  \\
      \hline                                                                   
      Simulation/Lepton Id     &      1.3\%  &  1.9\%   &  1.2\%  &       4.7\% &      11.3\% \\
      \hline
    \end{tabular}
    \caption{Sources of uncorrelated systematic uncertainties on  $\sigma_{\rm ZZ}$ and 
      $\sigma_{\rm ZZ\ra b\bar{b}X}$.}
  \end{center}
\end{table}

Three additional sources of systematic uncertainty, uncorrelated among
the channels, are considered:
the  
Monte Carlo statistics of the signal and the background  and the agreement between
data and 
Monte Carlo. The latter comprises  normalisation differences as derived from
the comparison of data and Monte Carlo samples five
to twenty times larger than the final ones, obtained by relaxing some
selection criteria. It also includes 
differences in the shape of the distribution of the lepton
identification variables
around the adopted selection requirements. All these uncertainties,
listed in Table~6,  do not affect the shape of the
discriminating distributions and their effect on the total cross
section propagates as summarised in Table~5.
The total systematic uncertainty is  the sum in quadrature
of all these contributions. The measured cross sections then read:
\begin{eqnarray*}
\sigma_{\rm ZZ}(192\GeV)\,\,\,  = & 0.29\pm 0.22\,{(\rm stat.)}\,{\pm
  0.02}\,{(\rm syst.)}\,{\rm pb}\,\,\,& (\rm SM: 0.79\pm0.02\,{\rm pb}),\\
\sigma_{\rm ZZ}(196\GeV)\,\,\,  = & 1.17\pm 0.24\,{(\rm stat.)}\,{\pm
  0.09}\,{(\rm syst.)}\,{\rm pb}\,\,\,& (\rm SM: 0.92\pm0.02\,{\rm pb}),\\
\sigma_{\rm ZZ}(200\GeV)\,\,\,  = & 1.25\pm 0.25\,{(\rm stat.)}\,{\pm
  0.09}\,{(\rm syst.)}\,{\rm pb}\,\,\,& (\rm SM: 1.00\pm0.02\,{\rm pb}),\\
\sigma_{\rm ZZ}(202\GeV)\,\,\,  = & 0.93\pm 0.38\,{(\rm stat.)}\,{\pm
  0.07}\,{(\rm syst.)}\,{\rm pb}\,\,\,& (\rm SM: 1.03\pm0.02\,{\rm pb}).
\end{eqnarray*}
The values in parentheses recall the Standard Model expectations.

A new fit is performed in terms of the NC02 framework and the corresponding cross
sections are derived as: 
\begin{eqnarray*}
\sigma_{\rm ZZ}^{\rm NC02}(192\GeV)\,\,\, =&0.29\pm 0.22\,{(\rm stat.)}\,{\pm
  0.02}\,{(\rm syst.)}\,{\rm pb}\,\,\,& (\rm SM: 0.77\pm0.02\,{\rm pb}),\\
\sigma_{\rm ZZ}^{\rm NC02}(196\GeV)\,\,\, =&1.18\pm 0.24\,{(\rm stat.)}\,{\pm
  0.09}\,{(\rm syst.)}\,{\rm pb}\,\,\,& (\rm SM: 0.90\pm0.02\,{\rm pb}),\\
\sigma_{\rm ZZ}^{\rm NC02}(200\GeV)\,\,\, =&1.25\pm 0.25\,{(\rm stat.)}\,{\pm
  0.09}\,{(\rm syst.)}\,{\rm pb}\,\,\,& (\rm SM: 0.98\pm0.02\,{\rm pb}),\\
\sigma_{\rm ZZ}^{\rm NC02}(202\GeV)\,\,\, =&0.95\pm 0.38\,{(\rm stat.)}\,{\pm
  0.07}\,{(\rm syst.)}\,{\rm pb}\,\,\,& (\rm SM: 1.01\pm0.02\,{\rm pb}).
\end{eqnarray*}
The Standard Model expectations given in parentheses are
calculated with the ZZTO~\cite{ZZTO} program and are assigned a 
$\pm2\%$ uncertainty~\cite{ZZTO}.
The YFSZZ~\cite{YFSZZ} package
yields compatible estimations.
As the relative
weights of the different final states are set according to the Z boson
branching fractions into fermions, the 
systematic uncertainty no longer includes
the $\pm2\%$ due to their predictions. On the other hand, a 
$\pm2\%$ uncertainty is assigned to account for
possible effects due to the 
extrapolation to the NC02 framework of the efficiencies and
background estimations from the Monte Carlo simulations
described above.

\subsection{b Quark Content in ZZ Events}

Z pair-production with at least a Z decaying into a b quark
pair constitutes an interesting test bench of the
detector capabilities to observe the minimal or a  supersymmetric Higgs
boson. These would in fact be seen as events with heavy
particles decaying into a b quark pair, recoiling against a Z boson.
Moreover, for the Higgs mass ranges under current investigation at LEP, the
cross sections of these processes are similar.

The $\qqqq$ final state analysis presents a high
sensitivity to  final states containing b quarks, as shown in
Figure~\ref{fig:2}b for the ${\rm b\bar bq\bar{q}}$ response of the
neural network used to select the $\qqqq$ final states.

\begin{figure}[p]
\begin{center}
    \begin{tabular}{c}
      \mbox{\rotatebox{90}{
          \includegraphics[width=.75\figwidth]{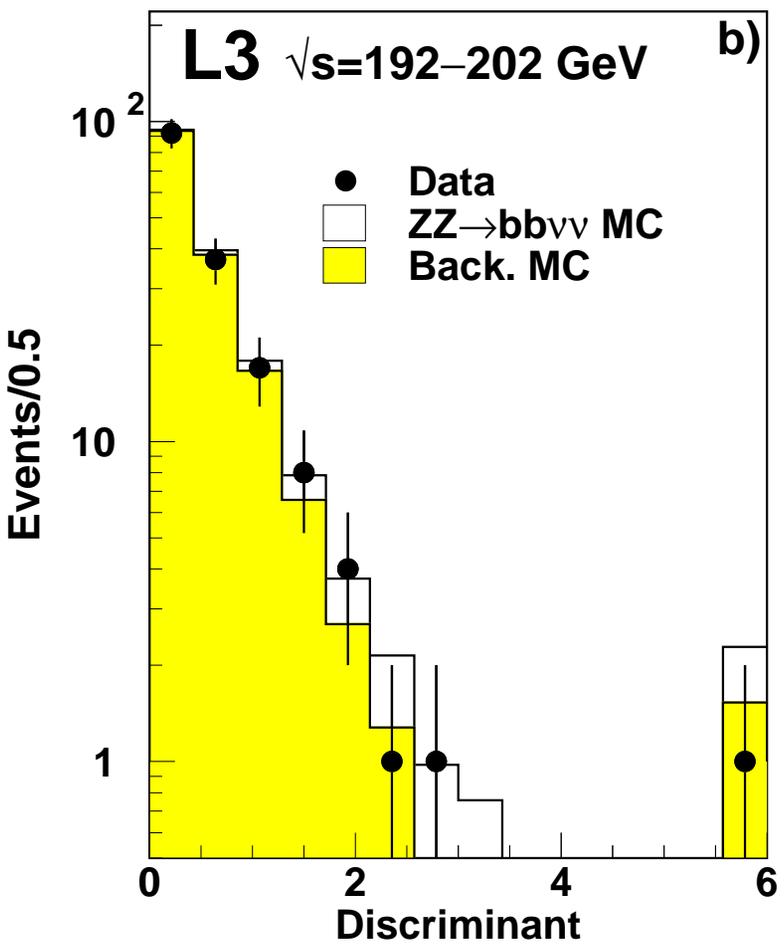}}} \\
      \mbox{\rotatebox{90}{
          \includegraphics[width=.75\figwidth]{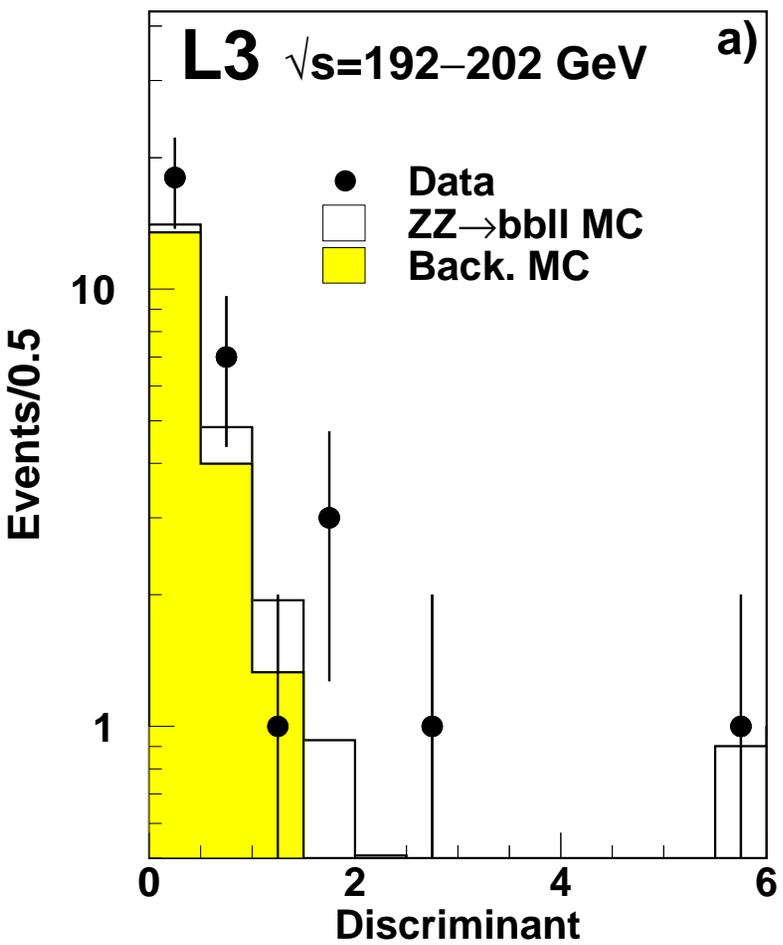}}} \\
    \end{tabular}
\caption{Discriminant variables in data and Monte Carlo for a) the
          $\rm b\bar{b}\ell^+\ell^-$ 
  and b) the $\rm b\bar{b}\nu\bar{\nu}$  selections. The last bin
  groups the overflows.}
\label{fig:4}
\end{center}
\end{figure}

The $\qqnn$ and
$\qqll$ selections,
as summarised by the distributions of ${ M_{5C}}$ for $\qqll$ and the
neural network 
output for $\qqnn$, are complemented with the same b-tag variable as
the $\qqqq$ 
selection. Its value is recorded for 
selected data and Monte Carlo events  for each of the two
hadronic jets. A single discriminant is then built for each channel
from its two b-tag variables and the selection one. First the variables are 
mapped to achieve uniform distributions for the background. Then the
product of their observed values is calculated event by event. Finally
the 
confidence level is calculated for the product of three uniformly distributed
quantities to be less than the observed product.
This confidence level is expected to be low for signal and flat
for background. The final discriminant is the negative logarithm
of this confidence level and is presented in Figure~\ref{fig:4}.

The cross section calculation for the individual channels is performed
as for the inclusive modes, considering the distributions in
Figures~\ref{fig:2}b and~\ref{fig:4}, and yields the results listed in 
Table~7.
The combined result for 
 $\sigma_{\rm ZZ\ra b\bar{b}X}$ is: 
\begin{displaymath}
\sigma_{\rm ZZ\ra b\bar{b}X}(192-202\GeV)=0.31 \pm 0.09\,{\rm
  (stat.)}\,\pm 0.03\,{\rm (syst.)}\,\mathrm{pb},
\end{displaymath}
in agreement with the Standard Model expectation of $0.27\pm0.01$\,pb. 
In all fits, the contribution from other Z pair final states are fixed
to their Standard Model 
expectations. The systematic uncertainties are evaluated in the same way as for
the total cross section, and are presented in Tables~5 and~6.

\begin{table}[ht]
  \begin{center}
    \begin{tabular}{|c|c|c|c|}
      \hline
      \rule{0pt}{12pt}  & $\rm b\bar{b}\ell^+\ell^-$ & $\rm b\bar{b}\nu\bar{\nu}$ &
      $\rm q \bar q b\bar{b}$ \\
      \hline
      \rule{0pt}{12pt} Measured cross section (pb)&
      $0.07 \pm 0.04$ &
      $<0.08$ &
      $0.26 \pm 0.07$ \\
      \rule{0pt}{12pt} Expected cross section (pb)&
      0.031 &
      0.057 &
      0.178 \\
      \hline
    \end{tabular}
    \caption{Results of the individual ZZ$\rm \ra b\bar{b}X$ cross
      section fits. The limit is at 95\% confidence level.}
  \end{center}
\end{table}

\begin{figure}[p]
\begin{center}
\includegraphics[width=15.0truecm]{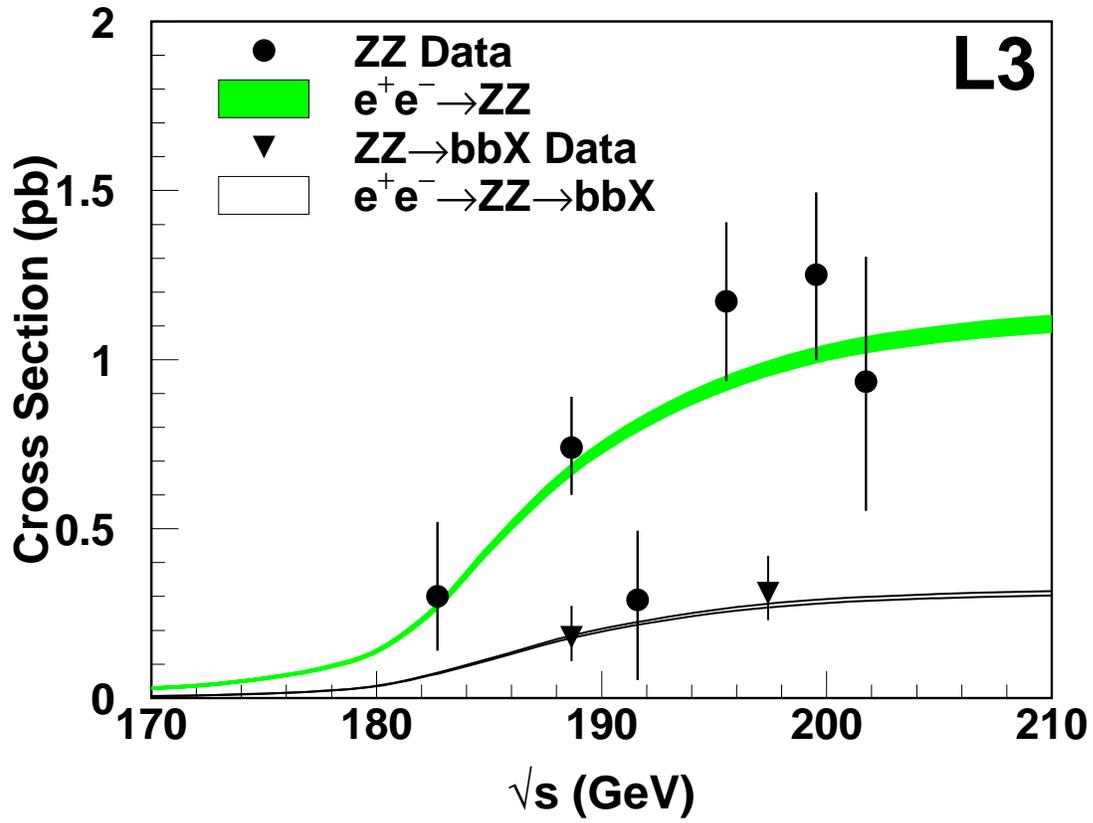}
\caption{Measurements of the $\rm\eeto ZZ$ and $\rm ZZ\ra
  b\bar{b}X$ cross sections, where statistical and systematic
  uncertainties are combined in quadrature.} 
\label{fig:5}
\end{center}
\end{figure}

Figure~\ref{fig:5} displays the measured total and ${\rm b\bar{b}X}$ cross
sections and their expected 
evolution with $\sqrt{s}$, including data at lower centre-of-mass
energies~\cite{zzl3183,l3zz189} and the theory uncertainties discussed above.

%
%
\section*{Acknowledgements}

We thank the CERN accelerator divisions for the
 continuous and successful upgrade of the
LEP machine and its excellent performance.  
We acknowledge the contributions of the engineers  and technicians who
have participated in the construction and maintenance of this experiment.


%
%

\newpage

\typeout{   }     
\typeout{Using author list for paper 221 ? }
\typeout{$Modified: Tue Sep  5 19:05:06 2000 by clare $}
\typeout{!!!!  This should only be used with paper 222 !!!!}
\typeout{!!!!  This should only be used with paper 222 !!!!}
\typeout{!!!!  This should only be used with paper 222 !!!!}
\typeout{!!!!  This should only be used with paper 222 !!!!}
\typeout{!!!!  This should only be used with paper 222 !!!!}
\typeout{!!!!  This should only be used with paper 222 !!!!}
\typeout{!!!!  This should only be used with paper 222 !!!!}
\typeout{   }
%
%
%
%
%
%

\newcount\tutecount  \tutecount=0
\def\tutenum#1{\global\advance\tutecount by 1 \xdef#1{\the\tutecount}}
\def\tute#1{$^{#1}$}
\tutenum\aachen            
\tutenum\nikhef            
\tutenum\mich              
\tutenum\lapp              
\tutenum\basel             
\tutenum\lsu               
\tutenum\beijing           
\tutenum\berlin            
\tutenum\bologna           
\tutenum\tata              
\tutenum\ne                
\tutenum\bucharest         
\tutenum\budapest          
\tutenum\mit               
\tutenum\debrecen          
\tutenum\florence          
\tutenum\cern              
\tutenum\wl                
\tutenum\geneva            
\tutenum\hefei             
\tutenum\seft              
\tutenum\lausanne          
\tutenum\lecce             
\tutenum\lyon              
\tutenum\madrid            
\tutenum\milan             
\tutenum\moscow            
\tutenum\naples            
\tutenum\cyprus            
\tutenum\nymegen           
\tutenum\caltech           
\tutenum\perugia           
\tutenum\cmu               
\tutenum\prince            
\tutenum\rome              
\tutenum\peters            
\tutenum\potenza           
\tutenum\salerno           
\tutenum\ucsd              
\tutenum\santiago          
\tutenum\sofia             
\tutenum\korea             
\tutenum\alabama           
\tutenum\utrecht           
\tutenum\purdue            
\tutenum\psinst            
\tutenum\zeuthen           
\tutenum\eth               
\tutenum\hamburg           
\tutenum\taiwan            
\tutenum\tsinghua          

{
\parskip=0pt
\noindent
{\bf The L3 Collaboration:}
\ifx\selectfont\undefined
 \baselineskip=10.8pt
 \baselineskip\baselinestretch\baselineskip
 \normalbaselineskip\baselineskip
 \ixpt
\else
 \fontsize{9}{10.8pt}\selectfont
\fi
\medskip
\tolerance=10000
\hbadness=5000
\raggedright
\hsize=162truemm\hoffset=0mm
\def\r{\rlap,}
\noindent

M.Acciarri\r\tute\milan\
P.Achard\r\tute\geneva\ 
O.Adriani\r\tute{\florence}\ 
M.Aguilar-Benitez\r\tute\madrid\ 
J.Alcaraz\r\tute\madrid\ 
G.Alemanni\r\tute\lausanne\
J.Allaby\r\tute\cern\
A.Aloisio\r\tute\naples\ 
M.G.Alviggi\r\tute\naples\
G.Ambrosi\r\tute\geneva\
% ...
\newpage
\begin{list}{A}{\itemsep=0pt plus 0pt minus 0pt\parsep=0pt plus 0pt minus 0pt
                \topsep=0pt plus 0pt minus 0pt}
\item[\aachen]
 I. Physikalisches Institut, RWTH, D-52056 Aachen, FRG$^{\S}$\\
 III. Physikalisches Institut, RWTH, D-52056 Aachen, FRG$^{\S}$
\item[\nikhef] National Institute for High Energy Physics, NIKHEF, 
     and University of Amsterdam, NL-1009 DB Amsterdam, The Netherlands
\item[\mich] University of Michigan, Ann Arbor, MI 48109, USA
\item[\lapp] Laboratoire d'Annecy-le-Vieux de Physique des Particules, 
     LAPP,IN2P3-CNRS, BP 110, F-74941 Annecy-le-Vieux CEDEX, France
\item[\basel] Institute of Physics, University of Basel, CH-4056 Basel,
     Switzerland
\item[\lsu] Louisiana State University, Baton Rouge, LA 70803, USA
\item[\beijing] Institute of High Energy Physics, IHEP, 
  100039 Beijing, China$^{\triangle}$ 
\item[\berlin] Humboldt University, D-10099 Berlin, FRG$^{\S}$
\item[\bologna] University of Bologna and INFN-Sezione di Bologna, 
     I-40126 Bologna, Italy
\item[\tata] Tata Institute of Fundamental Research, Bombay 400 005, India
\item[\ne] Northeastern University, Boston, MA 02115, USA
\item[\bucharest] Institute of Atomic Physics and University of Bucharest,
     R-76900 Bucharest, Romania
\item[\budapest] Central Research Institute for Physics of the 
     Hungarian Academy of Sciences, H-1525 Budapest 114, Hungary$^{\ddag}$
\item[\mit] Massachusetts Institute of Technology, Cambridge, MA 02139, USA
\item[\debrecen] KLTE-ATOMKI, H-4010 Debrecen, Hungary$^\P$
\item[\florence] INFN Sezione di Firenze and University of Florence, 
     I-50125 Florence, Italy
\item[\cern] European Laboratory for Particle Physics, CERN, 
     CH-1211 Geneva 23, Switzerland
\item[\wl] World Laboratory, FBLJA  Project, CH-1211 Geneva 23, Switzerland
\item[\geneva] University of Geneva, CH-1211 Geneva 4, Switzerland
\item[\hefei] Chinese University of Science and Technology, USTC,
      Hefei, Anhui 230 029, China$^{\triangle}$
\item[\seft] SEFT, Research Institute for High Energy Physics, P.O. Box 9,
      SF-00014 Helsinki, Finland
\item[\lausanne] University of Lausanne, CH-1015 Lausanne, Switzerland
\item[\lecce] INFN-Sezione di Lecce and Universit\`a Degli Studi di Lecce,
     I-73100 Lecce, Italy
\item[\lyon] Institut de Physique Nucl\'eaire de Lyon, 
     IN2P3-CNRS,Universit\'e Claude Bernard, 
     F-69622 Villeurbanne, France
\item[\madrid] Centro de Investigaciones Energ{\'e}ticas, 
     Medioambientales y Tecnolog{\'\i}cas, CIEMAT, E-28040 Madrid,
     Spain${\flat}$ 
\item[\milan] INFN-Sezione di Milano, I-20133 Milan, Italy
\item[\moscow] Institute of Theoretical and Experimental Physics, ITEP, 
     Moscow, Russia
\item[\naples] INFN-Sezione di Napoli and University of Naples, 
     I-80125 Naples, Italy
\item[\cyprus] Department of Natural Sciences, University of Cyprus,
     Nicosia, Cyprus
\item[\nymegen] University of Nijmegen and NIKHEF, 
     NL-6525 ED Nijmegen, The Netherlands
\item[\caltech] California Institute of Technology, Pasadena, CA 91125, USA
\item[\perugia] INFN-Sezione di Perugia and Universit\`a Degli 
     Studi di Perugia, I-06100 Perugia, Italy   
\item[\cmu] Carnegie Mellon University, Pittsburgh, PA 15213, USA
\item[\prince] Princeton University, Princeton, NJ 08544, USA
\item[\rome] INFN-Sezione di Roma and University of Rome, ``La Sapienza",
     I-00185 Rome, Italy
\item[\peters] Nuclear Physics Institute, St. Petersburg, Russia
\item[\potenza] INFN-Sezione di Napoli and University of Potenza, 
     I-85100 Potenza, Italy
\item[\salerno] University and INFN, Salerno, I-84100 Salerno, Italy
\item[\ucsd] University of California, San Diego, CA 92093, USA
\item[\santiago] Dept. de Fisica de Particulas Elementales, Univ. de Santiago,
     E-15706 Santiago de Compostela, Spain
\item[\sofia] Bulgarian Academy of Sciences, Central Lab.~of 
     Mechatronics and Instrumentation, BU-1113 Sofia, Bulgaria
\item[\korea]  Laboratory of High Energy Physics, 
     Kyungpook National University, 702-701 Taegu, Republic of Korea
\item[\alabama] University of Alabama, Tuscaloosa, AL 35486, USA
\item[\utrecht] Utrecht University and NIKHEF, NL-3584 CB Utrecht, 
     The Netherlands
\item[\purdue] Purdue University, West Lafayette, IN 47907, USA
\item[\psinst] Paul Scherrer Institut, PSI, CH-5232 Villigen, Switzerland
\item[\zeuthen] DESY, D-15738 Zeuthen, 
     FRG
\item[\eth] Eidgen\"ossische Technische Hochschule, ETH Z\"urich,
     CH-8093 Z\"urich, Switzerland
\item[\hamburg] University of Hamburg, D-22761 Hamburg, FRG
\item[\taiwan] National Central University, Chung-Li, Taiwan, China
\item[\tsinghua] Department of Physics, National Tsing Hua University,
      Taiwan, China
\item[\S]  Supported by the German Bundesministerium 
        f\"ur Bildung, Wissenschaft, Forschung und Technologie
\item[\ddag] Supported by the Hungarian OTKA fund under contract
numbers T019181, F023259 and T024011.
\item[\P] Also supported by the Hungarian OTKA fund under contract
  numbers T22238 and T026178.
\item[$\flat$] Supported also by the Comisi\'on Interministerial de Ciencia y 
        Tecnolog{\'\i}a.
\item[$\sharp$] Also supported by CONICET and Universidad Nacional de La Plata,
        CC 67, 1900 La Plata, Argentina.
\item[$\diamondsuit$] Also supported by Panjab University, Chandigarh-160014, 
        India.
\item[$\triangle$] Supported by the National Natural Science
  Foundation of China.
\end{list}
}
\vfill


\newpage

%
%

\clearpage
\newpage

\newpage

\newpage

\newpage

\newpage

\end{document}